\documentclass{article}
\usepackage{spconf,amsmath,graphicx}
\usepackage{amsfonts,amssymb}
\usepackage{float}
\usepackage{array}
\usepackage{booktabs}
\usepackage{epstopdf}
\usepackage{url}

\title{VISinger 2: High-fidelity End-to-end singing voice synthesis enhanced by digital signal processing synthesizer}
\name{Yongmao Zhang$^1$, Heyang Xue$^1$, Hanzhao Li$^1$, Lei Xie$^1$$^*$, Tingwei Guo$^2$, Ruixiong Zhang$^2$, Caixia Gong$^2$}
\address{
  $^1$Audio, Speech and Language Processing Group (ASLP@NPU)\\School of Computer Science,
  Northwestern Polytechnical University, Xi’an, China\\
  $^2$DiDi Chuxing, Beijing, China}

\begin{document}
\ninept
\maketitle
\begin{abstract}

\vspace{-2pt}
\renewcommand{\thefootnote}{\fnsymbol{footnote}}
\footnotetext{* Corresponding author.}

End-to-end singing voice synthesis~(SVS) model VISinger~\cite{ref_VISinger} can achieve better performance than the typical two-stage model with fewer parameters. However, VISinger has several problems: \textit{text-to-phase problem}, the end-to-end model learns the meaningless mapping of text-to-phase; \textit{glitches problem}, the harmonic components corresponding to the periodic signal of the voiced segment occurs a sudden change with audible artefacts; \textit{low sampling rate}, the sampling rate of 24KHz does not meet the application needs of high-fidelity generation with the full-band rate (44.1KHz or higher). In this paper, we propose VISinger~2 to address these issues by integrating the digital signal processing~(DSP) methods with VISinger. Specifically, inspired by recent advances in differentiable digital signal processing (DDSP)~\cite{ref_DDSP}, we incorporate a DSP synthesizer into the decoder to solve the above issues. The DSP synthesizer consists of a harmonic synthesizer and a noise synthesizer to generate periodic and aperiodic signals, respectively, from the latent representation z in VISinger. It supervises the posterior encoder to extract the latent representation without phase information and avoid the prior encoder modelling text-to-phase mapping. To avoid glitch artefacts, the HiFiGAN is modified to accept the waveforms generated by the DSP synthesizer as a condition to produce the singing voice. Moreover, with the improved waveform decoder, VISinger~2 manages to generate 44.1kHz singing audio with richer expression and better quality. Experiments on OpenCpop corpus~\cite{ref_opencpop} show that VISinger~2 outperforms VISinger, CpopSing and RefineSinger in both subjective and objective metrics. \renewcommand{\thefootnote}{\arabic{footnote}}Our audio samples are available on the demo website~\footnote[1]{Demo: \url{https://zhangyongmao.github.io/VISinger2/}}, and we will release our source code upon the acceptance of this paper.
% ~\footnote[2]{Code: \url{https://github.com/zhangyongmao/VISinger2}}. 

\end{abstract}

\vspace{-2pt}
\begin{keywords}
Singing voice synthesis, variational autoencoder, adversarial learning%, end-to-end
\end{keywords}

\vspace{-8pt}
\section{Introduction}
\label{sec:intro}
\vspace{-7pt}

% Singing voice synthesis~(SVS) is a task that generates singing voices from
Singing voice synthesis~(SVS) is a task that generates singing voices from the given music score and lyrics like human singers. Deep learning based SVS approaches~\cite{ren2020deepsinger,blaauw2020sequence,hono2018recent,gu2021bytesing,ref_xiaoicesing,ref_diffsinger} have attracted tremendous attention in recent years for their extraordinary performances and wide applications. Similar to text-to-speech (TTS), most of these SVS systems consist of two stages, the acoustic model first generates low-dimensional spectral representations of vocal signals, typically mel-spectrogram, from the music score and lyrics, and the vocoder subsequently converts these intermediate representations into the singing waveform. Although these systems achieve decent performances, the two-stage models are separately trained, and the human-crafted intermediate representations, such as the mel-spectrogram, may limit the expressiveness of the synthesized singing voice. 
% typically mel-spectrogram ,  highly expressive

We have recently proposed VISinger~\cite{ref_VISinger} -- an end-to-end (E2E) learned SVS approach based on VITS~\cite{ref_VITS} to mitigate the problems of two-stage systems. Specifically, VITS adopts the structure of CVAE to realize end-to-end speech synthesis. The posterior encoder extracts the latent representation $z$ from the linear spectrum, the decoder restores $z$ to the waveform, and the prior encoder provides a prior constraint for z according to the text. To better model singing, VISinger provides $z$ with more accurate frame level prior constraints under the guidance of F0 and provides extra prior information for the duration predictor. VISinger achieves superior performance over the typical two-stage systems such as Fastspeech~\cite{ref_fastspeech} + HiFiGAN~\cite{ref_HiFiGAN}.

Although VISinger advances the end-to-end SVS, it still has some drawbacks preventing its further application in real-world applications. First, the quality artefacts of the two-stage systems still exist in VISinger. Specifically, the audible glitches, such as spectral discontinuities and occasional mispronunciations, reduce the naturalness of the generated singing voice. Second, the sampling rate of the generated singing voice of VISinger is 24KHz, which does not meet the needs of high-fidelity (HiFi) applications which desire full-band audio (44.1KHz or higher).  

To address these inadequacies, we reanalyzed the architecture and components of the VISinger. The first and most significant issue is that the latent representation $z$ extracted by the posterior encoder may contain phase information due to the gradients passed back by the decoder when modelling the waveform. This could lead to mispronunciation because it is extremely challenging to predict the phase from the linguistic input reasonably. Secondly, the HiFiGAN~\cite{ref_HiFiGAN} architecture adopted in VISinger is not well designed for the SVS task. Its absence of modelling capabilities of rich variations on singing voice may lead to the glitches problem. Finally, a higher sampling rate SVS system relies on an improved decoder to provide better modelling capabilities.

\begin{figure*}[ht]
	\centering
	\includegraphics[scale=0.48]{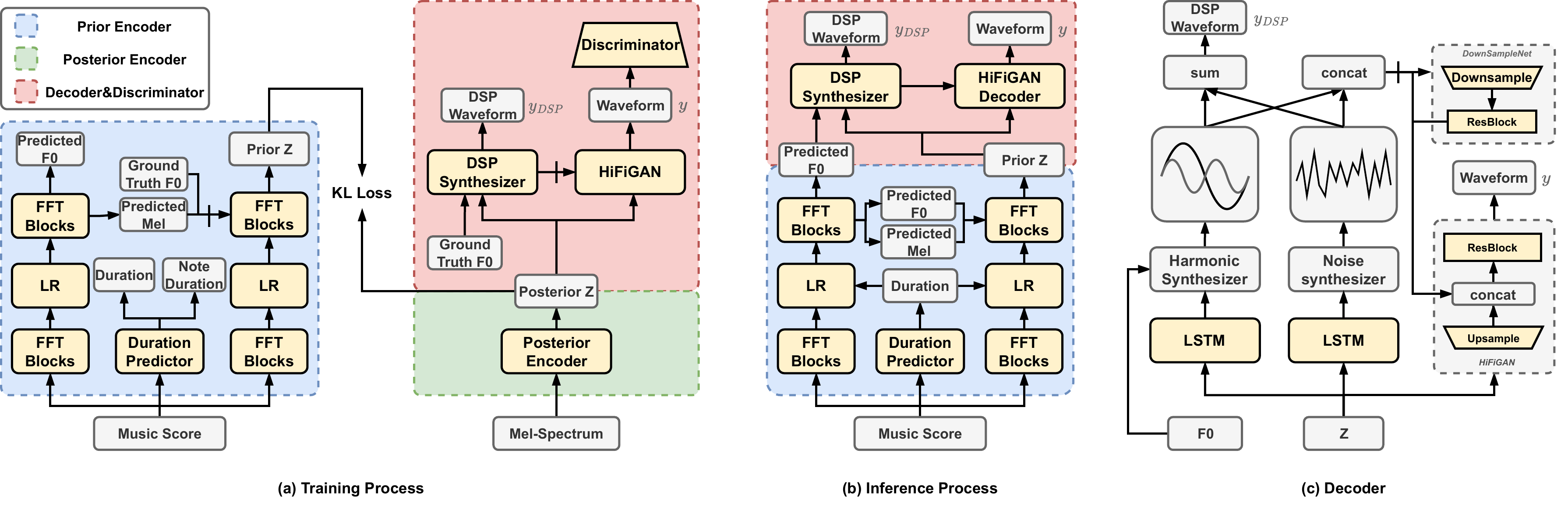}
	\vspace{-10pt}
	\caption{
		Architecture of VISinger~2. Yellow components are part of the neural network architecture, and grey components are features or differentiable operations. The short line on the arrow indicates gradient truncation.
	}
	\label{model}
	\vspace{-12pt}
\end{figure*}

In this paper, we propose VISinger~2, a digital signal processing~(DSP) synthesizer enhanced end-to-end SVS system for high-fidelity 44.1KHz singing generation. Specifically, inspired by recent advances in differentiable digital signal processing (DDSP)~\cite{ref_DDSP}, we incorporate a DSP synthesizer into VISinger to solve the above issues. Specifically, the DSP synthesizer consists of a harmonic synthesizer and a noise synthesizer to generate periodic and aperiodic signals from the latent representation $z$, respectively. The periodic and aperiodic signals are concatenated as conditional inputs to HiFiGAN, while the sum of the two produces a waveform to calculate the loss function. This design has sufficient advantages. First, both synthesizers need only amplitude information as input to generate the signals, thus fully compressing the phase component in $z$ and avoiding the text-to-phase challenge. Second, the representation of the periodic and aperiodic signal composition provides a strong condition for HiFi-GAN, substantially enhancing its modelling capability and allowing it to model a higher sampling rate. Finally, due to these improved modelling capabilities, the number of parameters in VISinger~2 can be substantially reduced by about 30\% compared to VISinger, further facilitating its use in real-world applications. Experiments show that VISinger~2 can generate a high-fidelity singing voice at a 44.1kHz sampling rate, with better naturalness and fewer glitches than VISinger and the traditional two-stage system.

We notice that there has been a recent trend to leverage the advances of conventional DSP to neural audio generation~\cite{ref_RefineGAN,ref_SingGAN,ref_Harm_SVC}. For example, in~\cite{ref_Harm_SVC}, harmonic signals are used to improve the stability of GAN and avoid pitch jitters and U/V errors in singing voice conversion. RefineGAN~\cite{ref_RefineGAN} calculates the speech template according to the pitch and then generates waveform according to the speech template. SingGAN~\cite{ref_SingGAN} adopts the source excitation with the adaptive feature learning filters to alleviate the glitch problem. These works usually focus on the periodic signal because the glitches problem comes from the defect of the periodic signal. Although motivated by these works aiming for better generation quality, our approach has substantial differences in terms of methodology. First, the above revisions are all made on vocoders, and the whole system still faces the two-stage mismatch problem. We mitigate this problem by proposing a fully end-to-end system VISinger~2. Second, to ensure that the extracted latent representation $z$ in VISinger~2 contains full amplitude information~(periodic and aperiodic parts), we leverage both periodic and aperiodic signals generated by the DSP synthesizer in our system design.

\vspace{-8pt}
\section{Method}
\label{sec:Method}
\vspace{-8pt}

The overall model architecture of VISinger~2 is shown in Fig.~\ref{model}. The proposed model adopts the conditional variational autoencoder (CVAE) structure, which includes three parts: a posterior encoder, a prior encoder and a decoder, the same as VITS~\cite{ref_VITS} and VISinger~\cite{ref_VISinger}. The posterior encoder extracts the latent representation~$z$ from spectral features, the decoder generates waveform~$y$ from $z$, and the prior conditional encoder constrains the extraction process of $z$. We will introduce the posterior encoder, decoder and prior encoder, respectively.

\vspace{-10pt}
\subsection{Posterior Encoder}
\label{sec2:Posterior Encoder}
\vspace{-5pt}

The posterior encoder is composed of multi-layer 1-D convolution, which aims to extract the latent representation~$z$ from the mel-spectrum. The last layer produces the mean and variance of the posterior distribution, and the resampling method is used to obtain the posterior $z$.

\vspace{-8pt}
\subsection{Decoder}
\label{sec2:Decoder}
\vspace{-4pt}

% HiFi-GAN\cite{ref_HiFiGAN} based 

The decoder generates waveform from the latent representation~$z$ as shown in Fig.\ref{model}(c). To avoid text-to-phase and glitches problems, we incorporate a DSP synthesizer into the decoder. Specifically, we use a harmonic synthesizer and a noise synthesizer to generate periodic and aperiodic parts of the waveform from the posterior $z$. The generated waveforms are used as an auxiliary condition for HiFi-GAN as input to enhance its modelling capabilities relieving the glitch problem. Meanwhile, since the inputs of both two synthesizers contain only amplitude information, the posterior $z$ will lean towards not including phase information and thus alleviate the text-to-phase problem. 

\vspace{-8pt}
\subsubsection{Harmonic Synthesizer}
\label{sec2:Periodic signal part}
\vspace{-5pt}

We use the harmonic synthesizer to generate harmonic components of audio the same as the harmonic oscillator in DDSP~\cite{ref_DDSP}. The harmonic synthesizer uses sin signals to simulate the waveform of each formant of the single sound source audio. The $k$-th sinusoidal component signal~$y_{k}$ generated by the harmonic synthesizer can be expressed as:
% , and uses the integral method to calculate the phase of sin function.

\vspace{-12pt}
\begin{equation}
  \setlength{\abovedisplayskip}{3pt}
  \setlength{\belowdisplayskip}{3pt}
\begin{split}
    y_{k}(n)=H_{k}(n)sin(\phi_{k}(n))
\end{split}
\end{equation}
where $n$ represents the time step of the sample sequence, and $H_{k}$ is the time-varying amplitude of the $k$-th sinusoidal component. The phase~$\phi_{k}(n)$ is obtained by integrating on the sample sequence:

\vspace{-3pt}
\begin{equation}
  \setlength{\abovedisplayskip}{3pt}
  \setlength{\belowdisplayskip}{3pt}
   \phi_{k}(n)=2\pi \sum_{m=0}^{n}\frac{f_{k}(m)}{Sr}+\phi_{0,k}
\end{equation}
where $f_{k}$ represents the frequency of the $k$-th sinusoidal component, $Sr$ represents the sampling rate, and $\phi_{0,k}$ represents the initial phase. We can get the phase of the sin signal~$y_{k}$ through an accumulation operation according to the fundamental frequency~$f_{k}$. The frequency~$f_{k}$ can be calculated by $f_{k}(n)=kf_{0}(n)$, where $f_{0}$ is the fundamental frequency. The time-varying $f_{k}$ and $H_{k}$ are interpolated from frame-level features. We extract the fundamental frequency using Harvest~\cite{ref_harvest} algorithm.

\vspace{-8pt}
\subsubsection{Noise Synthesizer}
\label{sec2:Aperiodic signal part}

\vspace{-5pt}
In the noise synthesizer, we use inverse short-time Fourier transform~(iSTFT) to generate the stochastic components of audio, similar to the filtered noise in DDSP. The aperiodic components are closer to noise, but the energy distribution is uneven in different frequency bands. The stochastic component signal~$y_{noise}$ generated can be expressed as:

\vspace{-13pt}
\begin{equation}
  \setlength{\abovedisplayskip}{3pt}
  \setlength{\belowdisplayskip}{3pt}
\begin{split}
    y_{noise}=iSTFT(N, P)
\end{split}
\end{equation}
where the phase spectrogram~$P$ of iSTFT is uniform noise in domain [$-\pi$, $\pi$], and the amplitude spectrogram~$N$ is predicted by the network.

\vspace{-5pt}
\subsubsection{Loss Function of Decoder}
\label{sec2:Combination}

\vspace{-5pt}
The DSP waveforms generated by the DSP synthesizer contain both harmonic and stochastic components. The complete DSP waveform $y_{DSP}$ and the loss~$L_{DSP}$ of the DSP synthesizer are defined as

\vspace{-10pt}
\begin{equation}
  \setlength{\abovedisplayskip}{3pt}
  \setlength{\belowdisplayskip}{3pt}
\begin{split}
    y_{DSP}=\sum_{k=0}^{K}y_{k}+y_{noise}
\end{split}
\end{equation}
\begin{equation}
\begin{split}
    L_{DSP}=\lambda_{DSP}\left\| \text{Mel}(y_{DSP}) - \text{Mel}(y)\right\|_{1}
\end{split}
\end{equation}
where $K$ represents the number of the sinusoidal component and $\text{Mel}$ represents the process of extracting mel-spectrum from waveform.

We use a downsampling network gradually downsamples the DSP waveforms to the frame-level features. The HiFi-GAN accepts the posterior z and the intermediate features generated by the downsampling network as input and generates the final waveform~$\hat{y}$. Following HiFi-GAN, the GAN loss for the generator G is defined as:

\vspace{-10pt}
\begin{equation}
  \setlength{\abovedisplayskip}{3pt}
  \setlength{\belowdisplayskip}{3pt}
\begin{split}
    L_{G}=L_{adv}(G)+\lambda_{fm} L{fm}+\lambda_{Mel}L_{Mel}
\end{split}
\end{equation}
where $L_{adv}$ is the adversarial loss, $L_{fm}$ is the feature matching loss, and $L_{Mel}$ is the Mel-Spectrogram loss. 

\vspace{-5pt}
\subsubsection{Discriminator}
\label{sec2:Discriminator}

\vspace{-5pt}
We combine two sets of discriminators to improve the ability of the discriminator. One set of discriminators is multi-resolution spectrogram discriminator~(MRSD) in UnviNet~\cite{ref_Univnet}, and the other is Multi-Period Discriminator~(MPD) and Multi-Scale Discriminator~(MSD) in HiFi-GAN~\cite{ref_HiFiGAN}. 

\vspace{-5pt}
\subsection{Prior Encoder}
\label{sec2:Prior Encoder}
\vspace{-5pt}

The prior encoder takes the music score as input to provide a prior constraint for CVAE. As mentioned in Section~\ref{sec2:Decoder}, the posterior $z$ will be used to predict $H$, $N$ in the decoder, where $H$ represents the amplitude of the sinusoidal formant and $N$ represents the amplitude spectrum of aperiodic components. Both $H$ and $N$ only contain amplitude information but not phase information, so the posterior $z$ will not contain phase information accordingly. In this way, the prior encoder will not model the text-to-phase mapping when predicting the posterior $z$ based on the music score.

Similar to VISinger~\cite{ref_VISinger}, the prior encoder adopts the same structure as Fastspeech~\cite{ref_fastspeech}. The flow~\cite{ref_flow} module plays an important role in VITS~\cite{ref_VITS}, but it occupies a large number of model parameters. For a more practical structure, we calculate the KL divergence $L_{kl}$ directly between the prior $z$ and the posterior $z$ without using flow. 

We use a separate FastSpeech~\cite{ref_fastspeech} model to predict the fundamental frequency and mel-spectrum to guide the frame-level prior networks. The loss for the auxiliary feature is defined as:

\vspace{-7pt}
\begin{equation}
  \setlength{\abovedisplayskip}{3pt}
  \setlength{\belowdisplayskip}{3pt}
\begin{split}
    L_{af}=\left\| LF0 - \widehat{LF0} \right\|_{2} + \left\| Mel - \widehat{Mel} \right\|_{1}
\end{split}
\vspace{-5pt}
\end{equation}
where $\widehat{LF0}$ is the predicted log-F0, and $\widehat{Mel}$ is the predicted mel-spectrogram. 

We take the predicted mel-spectrum as the auxiliary feature for the frame-level prior network in the training and inference process, so the auxiliary mel-spectrum does not bring a mismatch in the training and inference process. The frame-level prior network predicts the prior $z$ with the guide of auxiliary mel-spectrum to alleviate the text-to-phase problem further. We prove later in the experiment that VISinger~2 does not rely too much on this auxiliary mel-spectrum. The harmonic synthesizer accepts the predicted fundamental frequency as input to guide the generation of periodic signals in the inference process, while the ground-truth fundamental frequency is adopted in the training process.

The duration predictor accepts the music score as input and adopts the method in XiaoiceSing~\cite{ref_xiaoicesing} to simultaneously predict phoneme duration and note duration. The duration loss is expressed as:

\vspace{-13pt}
\begin{equation}
  \setlength{\abovedisplayskip}{3pt}
  \setlength{\belowdisplayskip}{3pt}
\begin{split}
    L_{dur}=\left\| d_{phone} - \widehat{d_{phone}} \right\|_{2} + \left\| d_{note} - \widehat{d_{note}} \right\|_{2}
\end{split}
\end{equation}
where $d_{phone}$ is the ground truth phoneme duration, $\widehat{d_{phone}}$ is the predicted phoneme duration, while $d_{note}$ is the ground truth note duration, and $\widehat{d_{note}}$ is the predicted note duration.

\vspace{-5pt}
\subsection{Final Loss}
\label{sec2:Final Loss}
\vspace{-5pt}
Our final objectives for the proposed model can be expressed as:

\vspace{-13pt}
\begin{equation}
  \setlength{\abovedisplayskip}{3pt}
  \setlength{\belowdisplayskip}{3pt}
\begin{split}
    L(G)=L_{G}+ L_{kl}+ L_{DSP} + L_{dur} + L_{af}
\end{split}
\end{equation}
\begin{equation}
  \setlength{\abovedisplayskip}{3pt}
  \setlength{\belowdisplayskip}{3pt}
    L(D)=L_{adv}(D)
\end{equation}
where $L_{G}$ is the GAN loss for generator G, $L_{kl}$ is KL divergence between prior~$z$ and posterior~$z$, $L_{af}$ is the loss of the auxiliary feature, and $L_{adv}(D)$ is the GAN loss of discriminator D.

\vspace{-5pt}
\section{Experiments}
\label{sec:exp}

\vspace{-8pt}
\subsection{Datasets}
\label{sec3:Datasets}
\vspace{-3pt}

We evaluate VISinger~2 on the Opencpop~\cite{ref_opencpop} dataset, which consists of 100 popular Mandarin songs~(5.2 hours) performed by a female professional singer. All the audios are recorded at 44.1kHz with 16-bit quantization. Opencpop has a pre-defined training set and test set: 3,550 segments from 95 songs for training while 206 segments from 5 songs for the test. We follow Opencpop's division of the training and test set.

\vspace{-7pt}
\subsection{Model Configuration}
\label{sec3:Model Configuration}
\vspace{-5pt}

We train the following systems for comparison.
\begin{itemize}
  \item
  \textbf{CpopSing}: the two-stage conformer-based SVS model introduced in the Opencpop~\cite{ref_opencpop}. In the CpopSing, the Transformer blocks in Fastspeech~2~\cite{ref_FastSpeech2} are replaced with Conformer blocks. The adversarial training method similar to the sub-frequency adversarial loss in HiFiSinger~\cite{ref_HiFiSinger} is used in the CpopSing.
  \item\vspace{-3pt}
  \textbf{VISinger}: an end-to-end SVS system based on VITS. The model configuration is consistent with that in VISinger~\cite{ref_VISinger}.
  \item\vspace{-3pt}
  \textbf{RefineSinger}: a two-stage SVS system constructed by FastSpeech~\cite{ref_fastspeech} and RefineGAN~\cite{ref_RefineGAN}. The FFT block in both the encoder and decoder of Fastspeech are 4-layer. The duration predictor consists of a 3-layer 1D-convolutional network and predicts the phoneme-level and note-level duration. RefineGAN, which is designed for high sampling rate scenarios, adopts pitch-guided architecture to improve the ability of the generator. A Mel2F0 module introduced in \cite{ref_learn2sing2} is used to predict the F0 for RefineGAN. The hidden dimension of RefineGAN is 512, and the data augmentation method proposed in \cite{ref_RefineGAN} is not employed for simplicity.
  \item\vspace{-3pt}
  \textbf{VISinger~2}: the proposed end-to-end SVS system, adopting all the contributions introduced in the paper. Each FFT blocks in VISinger2 consist of 4-layer FFTs. The hidden dim and filter dim of FFT are 192 and 768, respectively. The hidden dimension of HiFi-GAN in the decoder is 256. The posterior encoder consists of an 8-layer 1D-convolutional network, and the dimension of potential representation $z$ is 192. The duration predictor consists of a 3-layer 1D-convolutional network with ReLU activation.
\end{itemize}

All models are trained up to 500k steps with a batch size of 16. The Adam optimizer with $\beta_{1}$ = 0.8, $\beta_{2}$ = 0.99 and $\epsilon$ = $10^{-9}$ is used to train all the models.

\begin{table}[]
	\centering
	\caption{Experimental results in terms of subjective mean opinion score~(MOS) and two objective metrics.}
\begin{tabular}{lccccc}
\bottomrule
Model             & \makebox[0.05\textwidth][c]{\begin{tabular}[c]{@{}c@{}}Sample\\ Rate\end{tabular}}     & \makebox[0.04\textwidth][c]{\begin{tabular}[c]{@{}c@{}}Model\\ Size~(M)\end{tabular}} & \makebox[0.04\textwidth][c]{\begin{tabular}[c]{@{}c@{}}F0\\ RMSE\end{tabular}} & \makebox[0.04\textwidth][c]{\begin{tabular}[c]{@{}c@{}}Dur\\ RMSE\end{tabular}} & MOS \\ \hline
Cpopsing &22k     & 137.5                                                & 28.5                                              & 6.6                                                &  2.97$\pm$0.12   \\
VISinger &22k     & 36.5                                                 & 33.7                                              & 3.6                                                &  3.46$\pm$0.13 \\
VISinger~2 &22k    & \textbf{25.7}                                        & \textbf{26.0}                                     & 2.8                                                &  3.69 $\pm$0.15 \\\hline

RefineSinger &44k & 36.0                                                 & 39.1                                              & 2.8                                                &  2.85$\pm$0.10 \\

VISinger~2 &44k    & \textbf{25.7}                                        & 26.7                                              & \textbf{2.7}                                       &  \textbf{3.81$\pm$0.14}  \\ \hline
Recording &22k       & -                                                    & -                                                 & -                                                  &  4.22$\pm$0.12  \\
Recording &44k       & -                                                    & -                                                 & -                                                  &  4.32$\pm$0.11  \\ \bottomrule
\end{tabular}
\label{MOS}
\vspace{-10pt}
\end{table}

\vspace{-6pt}
\subsection{Experimental Results}
\label{sec3:Experimental}
\vspace{-6pt}

We performed a mean opinion score (MOS) test for the above systems and randomly selected 30 segments from the test set for subjective listening, and ten listeners attended the test. The objective metrics, including F0 Root Mean Square Error~(F0-RMSE) and duration Root Mean Square Error~(dur-RMSE), are calculated to evaluate the performance of different systems. The results are summarized in Table~\ref{MOS}. 

To evaluate the performance of the proposed VISinger~2 in a general SVS scenario, we first compared VISinger~2 with CpopSing and VISinger at the 22.05kHz sampling rate. As shown in Table~\ref{MOS}, VISinger~2 and VISinger perform significantly better than CpopSing in the MOS test, demonstrating the superiority of the end-to-end model in the general SVS scenario. Meanwhile, the MOS score of VISinger~2 is higher than VISinger by about 0.23, indicating the effectiveness of our design in a general SVS scenario. For further validation of the performance of VISinger2 in high sampling rate SVS scenarios, we compared VISinger~2 with RefineSinger at the 44.1 kHz sampling rate. The evaluation results listed in Table~\ref{MOS} show that VISinger~2 surpasses RefineSinger in MOS score by 33.6\% and has a MOS improvement of about 0.15 compared to the 22.05 kHz version of VISinger~2. This improvement shows that VISinger~2 is capable of modelling high sampling rates SVS enables high-fidelity singing voice generation. Note that CpopSing and VISinger did not participate in the 44.1kHz comparison for fairness as they are not designed for high sampling rate SVS. Similar to the MOS results, VISinger~2 outperformed the other systems in terms of objective metrics, validating our assumptions again. 

Another observation worth highlighting is that in addition to outperforming the other systems in MOS and objective metrics, VISinger~2 has the smallest number of parameters in all comparison systems at 25.7M. This result demonstrates the effectiveness of our proposed approach and its sufficiency to be applied in real-world scenarios.

% \vspace{-7pt}
\begin{figure}[h]
	\centering
	\includegraphics[width=0.98\linewidth]{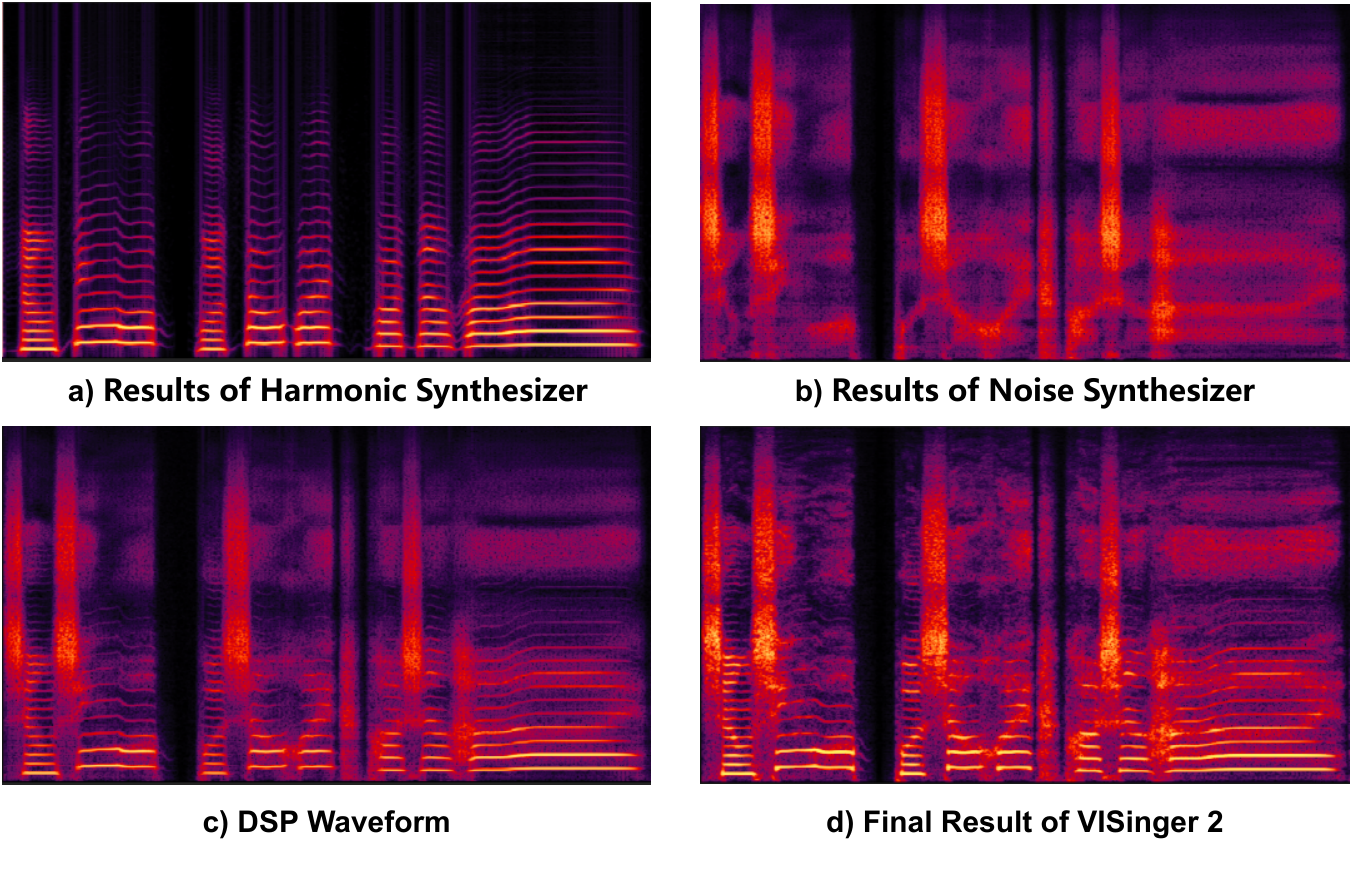}
    \vspace{-12pt}
	\caption{
		Visualization of synthesized waveform.
	}
    \vspace{-15pt}
	\label{exp_spec}
\end{figure}

% \vspace{-8pt}
We further visualize the waveforms generated by VISinger~2 in Fig.~\ref{exp_spec} to illustrate the role of the DSP synthesizer. As shown in Fig.~\ref{exp_spec}, the periodic components and aperiodic components are generated by the harmonic synthesizer and noise synthesizer, respectively. The generated periodic and aperiodic components are added to get DSP waveform~$y_{DSP}$. We can also find that the waveform finally generated by HiFi-GAN is guided by the DSP waveform as its conditional input.%Finally, the HiFi-GAN accepts the harmonic and the aperiodic components as a condition to produce the final waveform~$\hat{y}$.

\vspace{-13pt}
\begin{table}[H]
\centering
	\caption{Ablation study results in terms of subjective mean opinion score~(MOS).}
\begin{tabular}{lcc}
\hline
\multicolumn{1}{c}{Model} &\begin{tabular}[c]{@{}c@{}}Sample\\ Rate\end{tabular} & MOS  \\ \hline
Recording &44k              & 4.47$\pm$0.09 \\
VISinger2&44k            & 3.96$\pm$0.11 \\
\ \ -auxiliary mel-spectrum&44k                      & 3.85$\pm$0.12 \\
\ \ -DSP synthesizer    &44k      & 3.02$\pm$0.13 \\ \hline
\end{tabular}\vspace{-7pt}
	\label{MOS_ab}
\end{table}

\vspace{-12pt}
\subsection{Ablation study}
\label{sec3:Ablation study}
\vspace{-3pt}

To validate the effectiveness of each contribution, we conduct an ablation study. We remove the DSP synthesizer and auxiliary mel-spectrum feature, respectively. The results are summarized in Table~\ref{MOS_ab}. The results show that the model's performance degrades significantly when the DSP synthesizer is deleted, indicating that the DSP synthesizer plays an essential role in solving the text-to-phase problem and glitches problems. At the same time, when the auxiliary mel-spectrum feature is deleted, the model's performance degrades slightly, indicating that the auxiliary mel-spectrum can further solve the text-to-phase problem because a complete mel-spectrum guides the prediction of the prior $z$.
% indicating that the auxiliary feature only plays an auxiliary role in solving the text-to-phase problem, and the VISinger~2 does not rely too much on the auxiliary feature. 

\vspace{-2pt}
\section{Conclusions}
\label{sec:Conclusions}
\vspace{-6pt}

In this work, we have updated our previous end-to-end singing voice synthesis system VISinger to its new version VISinger~2. Specifically, we solved the text-to-phase problem and the glitch artefacts problem and upgraded the sampling rate from 24KHz to 44.1KHz for a high-fidelity singing generation. These new contributions were achieved by incorporating a differential digital signal processing (DDSP) synthesizer with the VISinger decoder. In this way, the posterior encoder extracts the latent representation without phase information and avoids the prior encoder modelling text-to-phase mapping. To avoid glitch artefacts, we modified the decoder to accept the waveforms generated by the DSP synthesizer as a condition to produce the singing voice. Our experimental results show that, with fewer model parameters, VISinger~2 substantially outperforms CpopSing, VISinger and RefineSinger.

\vfill\pagebreak

\bibliographystyle{IEEE}
\bibliography{strings,refs}

\begin{thebibliography}{10}

\bibitem{ref_VISinger}
Yongmao Zhang, Jian Cong, Heyang Xue, Lei Xie, Pengcheng Zhu, and Mengxiao Bi,
\newblock ``Visinger: Variational inference with adversarial learning for
  end-to-end singing voice synthesis,''
\newblock in {\em {IEEE} International Conference on Acoustics, Speech and
  Signal Processing, {ICASSP} 2022, Virtual and Singapore, 23-27 May 2022}.
  2022, pp. 7237--7241, {IEEE}.

\bibitem{ref_DDSP}
Jesse~H. Engel, Lamtharn Hantrakul, Chenjie Gu, and Adam Roberts,
\newblock ``{DDSP:} differentiable digital signal processing,''
\newblock in {\em 8th International Conference on Learning Representations,
  {ICLR} 2020, Addis Ababa, Ethiopia, April 26-30, 2020}. 2020, OpenReview.net.

\bibitem{ref_opencpop}
Yu~Wang, Xinsheng Wang, Pengcheng Zhu, Jie Wu, Hanzhao Li, Heyang Xue, Yongmao
  Zhang, Lei Xie, and Mengxiao Bi,
\newblock ``Opencpop: {A} high-quality open source chinese popular song corpus
  for singing voice synthesis,''
\newblock in {\em Interspeech 2022, 23rd Annual Conference of the International
  Speech Communication Association, Incheon, Korea, 18-22 September 2022}.
  2022, pp. 4242--4246, {ISCA}.

\bibitem{ren2020deepsinger}
Yi~Ren, Xu~Tan, Tao Qin, Jian Luan, Zhou Zhao, and Tie-Yan Liu,
\newblock ``Deepsinger: Singing voice synthesis with data mined from the web,''
\newblock in {\em Proceedings of the 26th ACM SIGKDD International Conference
  on Knowledge Discovery \& Data Mining}, 2020.

\bibitem{blaauw2020sequence}
Merlijn Blaauw and Jordi Bonada,
\newblock ``Sequence-to-sequence singing synthesis using the feed-forward
  transformer,''
\newblock in {\em 2020 IEEE International Conference on Acoustics, Speech and
  Signal Processing (ICASSP)}, 2020.

\bibitem{hono2018recent}
Yukiya Hono, Shumma Murata, Kazuhiro Nakamura, Kei Hashimoto, Keiichiro Oura,
  Yoshihiko Nankaku, and Keiichi Tokuda,
\newblock ``Recent development of the dnn-based singing voice synthesis
  system—sinsy,''
\newblock in {\em Proceedings, APSIPA Annual Summit and Conference}, 2018.

\bibitem{gu2021bytesing}
Yu~Gu, Xiang Yin, Yonghui Rao, Yuan Wan, Benlai Tang, Yang Zhang, Jitong Chen,
  Yuxuan Wang, and Zejun Ma,
\newblock ``Bytesing: A chinese singing voice synthesis system using duration
  allocated encoder-decoder acoustic models and wavernn vocoders,''
\newblock in {\em 2021 12th International Symposium on Chinese Spoken Language
  Processing (ISCSLP)}, 2021.

\bibitem{ref_xiaoicesing}
Peiling Lu, Jie Wu, Jian Luan, Xu~Tan, and Li~Zhou,
\newblock ``Xiaoicesing: {A} high-quality and integrated singing voice
  synthesis system,''
\newblock in {\em Interspeech 2020, 21st Annual Conference of the International
  Speech Communication Association, Virtual Event, Shanghai, China, 25-29
  October 2020}. 2020, pp. 1306--1310, {ISCA}.

\bibitem{ref_diffsinger}
Jinglin Liu, Chengxi Li, Yi~Ren, Feiyang Chen, and Zhou Zhao,
\newblock ``Diffsinger: Singing voice synthesis via shallow diffusion
  mechanism,''
\newblock in {\em Thirty-Sixth {AAAI} Conference on Artificial Intelligence,
  {AAAI} 2022, Thirty-Fourth Conference on Innovative Applications of
  Artificial Intelligence, {IAAI} 2022, The Twelveth Symposium on Educational
  Advances in Artificial Intelligence, {EAAI} 2022 Virtual Event, February 22 -
  March 1, 2022}. 2022, pp. 11020--11028, {AAAI} Press.

\bibitem{ref_VITS}
Jaehyeon Kim, Jungil Kong, and Juhee Son,
\newblock ``Conditional variational autoencoder with adversarial learning for
  end-to-end text-to-speech,''
\newblock in {\em Proceedings of the 38th International Conference on Machine
  Learning, {ICML} 2021, 18-24 July 2021, Virtual Event}. 2021, vol. 139 of
  {\em Proceedings of Machine Learning Research}, pp. 5530--5540, {PMLR}.

\bibitem{ref_fastspeech}
Yi~Ren, Yangjun Ruan, Xu~Tan, Tao Qin, Sheng Zhao, Zhou Zhao, and Tie{-}Yan
  Liu,
\newblock ``Fastspeech: Fast, robust and controllable text to speech,''
\newblock in {\em Advances in Neural Information Processing Systems 32: Annual
  Conference on Neural Information Processing Systems 2019, NeurIPS 2019,
  December 8-14, 2019, Vancouver, BC, Canada}, 2019, pp. 3165--3174.

\bibitem{ref_HiFiGAN}
Jungil Kong, Jaehyeon Kim, and Jaekyoung Bae,
\newblock ``Hifi-gan: Generative adversarial networks for efficient and high
  fidelity speech synthesis,''
\newblock in {\em Advances in Neural Information Processing Systems 33: Annual
  Conference on Neural Information Processing Systems 2020, NeurIPS 2020,
  December 6-12, 2020, virtual}, 2020.

\bibitem{ref_RefineGAN}
Shengyuan Xu, Wenxiao Zhao, and Jing Guo,
\newblock ``Refinegan: Universally generating waveform better than ground truth
  with highly accurate pitch and intensity responses,''
\newblock in {\em Interspeech 2022, 23rd Annual Conference of the International
  Speech Communication Association, Incheon, Korea, 18-22 September 2022}.
  2022, pp. 1591--1595, {ISCA}.

\bibitem{ref_SingGAN}
Feiyang Chen, Rongjie Huang, Chenye Cui, Yi~Ren, Jinglin Liu, Zhou Zhao,
  Nicholas~Jing Yuan, and Baoxing Huai,
\newblock ``Singgan: Generative adversarial network for high-fidelity singing
  voice generation,''
\newblock {\em CoRR}, vol. abs/2110.07468, 2021.

\bibitem{ref_Harm_SVC}
Haohan Guo, Zhiping Zhou, Fanbo Meng, and Kai Liu,
\newblock ``Improving adversarial waveform generation based singing voice
  conversion with harmonic signals,''
\newblock in {\em {IEEE} International Conference on Acoustics, Speech and
  Signal Processing, {ICASSP} 2022, Virtual and Singapore, 23-27 May 2022}.
  2022, pp. 6657--6661, {IEEE}.

\bibitem{ref_harvest}
Masanori Morise,
\newblock ``Harvest: {A} high-performance fundamental frequency estimator from
  speech signals,''
\newblock in {\em Interspeech 2017, 18th Annual Conference of the International
  Speech Communication Association, Stockholm, Sweden, August 20-24, 2017},
  Francisco Lacerda, Ed. 2017, pp. 2321--2325, {ISCA}.

\bibitem{ref_Univnet}
Won Jang, Dan Lim, Jaesam Yoon, Bongwan Kim, and Juntae Kim,
\newblock ``Univnet: {A} neural vocoder with multi-resolution spectrogram
  discriminators for high-fidelity waveform generation,''
\newblock in {\em Interspeech 2021, 22nd Annual Conference of the International
  Speech Communication Association, Brno, Czechia, 30 August - 3 September
  2021}. 2021, pp. 2207--2211, {ISCA}.

\bibitem{ref_flow}
Laurent Dinh, Jascha Sohl{-}Dickstein, and Samy Bengio,
\newblock ``Density estimation using real {NVP},''
\newblock in {\em 5th International Conference on Learning Representations,
  {ICLR} 2017, Toulon, France, April 24-26, 2017, Conference Track
  Proceedings}. 2017, OpenReview.net.

\bibitem{ref_FastSpeech2}
Yi~Ren, Chenxu Hu, Xu~Tan, Tao Qin, Sheng Zhao, Zhou Zhao, and Tie{-}Yan Liu,
\newblock ``Fastspeech 2: Fast and high-quality end-to-end text to speech,''
\newblock in {\em 9th International Conference on Learning Representations,
  {ICLR} 2021, Virtual Event, Austria, May 3-7, 2021}. 2021, OpenReview.net.

\bibitem{ref_HiFiSinger}
Jiawei Chen, Xu~Tan, Jian Luan, Tao Qin, and Tie{-}Yan Liu,
\newblock ``Hifisinger: Towards high-fidelity neural singing voice synthesis,''
\newblock {\em CoRR}, vol. abs/2009.01776, 2020.

\bibitem{ref_learn2sing2}
Heyang Xue, Xinsheng Wang, Yongmao Zhang, Lei Xie, Pengcheng Zhu, and Mengxiao
  Bi,
\newblock ``Learn2sing 2.0: Diffusion and mutual information-based target
  speaker {SVS} by learning from singing teacher,''
\newblock in {\em Interspeech 2022, 23rd Annual Conference of the International
  Speech Communication Association, Incheon, Korea, 18-22 September 2022}.
  2022, pp. 4267--4271, {ISCA}.

\end{thebibliography}

\end{document}